# A Shocking Solar Nebula?

Kurt Liffman

CSIRO/MSE, P.O. Box 56, Highett, Victoria 3190, AUSTRALIA

**ABSTRACT**

It has been suggested that shock waves in the solar nebula formed the high temperature materials observed in meteorites and comets. It is shown that the temperatures at the inner rim of the solar nebula could have been high enough over a sufficient length of time to produce chondrules, CAIs, refractory dust grains and other high-temperature materials observed in comets and meteorites.

The solar bipolar jet flow may have produced an enrichment of $^{16}$O in the solar nebula over time and the chondrule oxygen isotopic reservoirs are possibly due to temporal changes in the relative position of the inner edge of the solar nebula and the subsequent strength of the solar bipolar jet flow. As such, nebula heating models, such as the shock model, are not required to explain the formation of most high-temperature chondritic components.

*Subject Headings:* accretion disks – comets: general – solar system: formation – meteoroids

1. Introduction

In 2004, the Stardust mission sampled dust particles from Comet Wild 2 - a comet that was probably formed in the outer regions of the solar system, possibly in the Kuiper Belt region some 30 to 50 Astronomical Units from the Sun (Brownlee et al. 2006). The latest results from the Stardust mission (Nakamura et al. 2008), show that some of the particles from comet Wild 2 are similar to the chondrules found in primitive meteorites. These results are consistent with the predictions made by the Jet Flow (Liffman and Brown 1996) and X-wind models (Shu et al. 1996)[1], which suggest that the possible building blocks of the planets: chondrules, calcium aluminium inclusions (CAIs) and other refractory meteoritic materials, were formed close to the Sun and then ejected, via the agency of a bipolar jet flow, to other regions of the solar system (Skinner 1990, Liffman 1992).

Despite this evidence, a section of the planetary science community hold to the view that shock waves, in the outer regions of the solar nebula, processed solar nebula dust and produced the observed high temperature, meteoritic/cometary materials (Desch 2007, Bouwman 2008). In this study, it is suggested that the temperatures in the inner-most regions of the solar nebula were high enough for a sufficient length of time to explain the formation of the high temperature materials found in both meteorites and comets.

---

[1] The formation mechanism for bipolar jet flows is still not understood, as such, we prefer the general term "Jet Flow". The "X-wind" model is a specific model of bipolar jet flow formation, which may or may not be valid.

2. Heating

In the shock wave model, the peak temperature of a particle subjected to a shock wave is proportional to $(\rho v^3)^{1/4}$ (Liffman & Toscano 2000), where $\rho$ is the gas density and $v$ is the speed of the gas relative to the particle. Within the solar nebula, the values of $\rho$ and $v$ may have varied over many orders of magnitude (e.g., $0 - \sim 10^{-5}$ g cm$^{-3}$ for $\rho$, $0 - \sim 100$ km s$^{-1}$ for $v$), so the peak shock temperature could have been highly variable. The meteoritic record, however, tells us that chondrules were consistently formed by heating fine-grained material to temperatures in the range of 1700 to 2100 K on a timescale of millions of years (Scott 2007). It is unclear how shock waves could have restricted particle processing temperatures to this range over such a timescale.

The heating mechanisms in the Jet Flow Model are direct solar radiation from the Sun and diffuse radiation from the inner rim of the solar nebula (Vinković et al. 2006). In this model, chondrules, CAIs and other refractory meteoritic/cometary materials were formed in a jet flow which was produced at the inner edge of the solar nebula (Liffman 2007). They were then ejected, by the jet flow, to other regions of the nebula. As the particles moved away from the disk, their exposure to the radiation field from the inner disk decreased and their temperature dropped to that produced from mainly direct solar radiation. A schematic summary of the structure of the inner solar nebula is given in Figure 1

To understand the heating at the inner rim of the solar nebula, it is necessary to obtain an estimate of the distance between the Sun and the inner edge of the nebula as a function of time. In Figure 1, $R_t$ is the distance of the inner edge of the disk from the center of a star, where the stellar magnetosphere truncates the accretion disk. From Frank et al. (2002)

$$R_t \approx \left( \frac{4\pi}{\mu_0} \frac{B_*^2 R_*^6}{\dot{M}_a \sqrt{GM_*}} \right)^{2/7} = 0.067 \left( \frac{(B_*(R_*)/0.1\,\text{T})^2 (R_*/2R_\odot)^6}{(\dot{M}_a/10^{-8}\,\text{M}_\odot\,\text{year}^{-1})(M_*/M_\odot)^{1/2}} \right)^{2/7} \text{AU} \quad (1)$$

where $\mu_0$ is the permeability of free space, $B_*$ is the stellar magnetic field strength, $R_*$ is the stellar radius, $\dot{M}_a$ is the accretion rate of material onto the star, $G$ the universal gravitational constant and $M_*$ the mass of the star (in this study we have fixed $M_* = M_\odot$). We assume that the mass accretion rate onto the star has the form:

$$\dot{M}_a(t) \approx \dot{M}_a(t_0) \left( \frac{t}{t_0} \right)^{-\eta}, \quad (2)$$

with $t > 10^4$ year, $t_0 = 10^6$ year, $\eta = 1.5$ and $\dot{M}_a(t_0) \approx 4 \times 10^{-8}\,\text{M}_\odot\,/\,\text{year}$ (Hartmann et al. 1998).

Another relevant length-scale is the co-rotation radius, $R_{co}$, which is the distance from the star where the angular rotation frequency of the star is equal to the angular frequency of the accretion disk surrounding the star:

$$R_{co} = \left( \frac{GM_*}{\Omega_*^2} \right)^{1/3} = 0.078 \left( \left( \frac{M_*}{M_\odot} \right) \left( \frac{P_*}{8\,\text{days}} \right)^2 \right)^{1/3} \text{AU}, \quad (3)$$

where $\Omega_*$ is the angular rotational frequency of the star.

We can use the values for the luminosity and radius of the early Sun as obtained from a standard solar model (Figure 2) and work out the temperature of a particle at the inner edge of the solar nebula. Where, the particle temperature $T_p$ is described by the equation (Vinković et al. 2006)

$$T_p \approx \left(3\frac{(L_* + L_a)}{16\pi\sigma r^2}\right)^{1/4} = 1645\left(\frac{((L_* + L_a)/L_\odot)}{(r/0.05 \text{ AU})^2}\right)^{1/4} \text{ K}, \qquad (4)$$

where $L_*$ is the luminosity of the star, $L_a$ is the accretion luminosity ($L_a \approx \frac{G\dot{M}_a M_*}{R_*}$), $\sigma$ is the Stefan-Boltzmann constant, $r$ is the distance from the star (in this case $r = R_t$), and we have assumed that the particle absorption and emissivity factors are approximately equal to one.

The results of these calculations are shown in Figure 3. As can be seen, at around $10^5$ years, the particle temperatures are close to the CAI formation temperatures (Scott 2007). Between $10^5$ and $\sim 2\times10^6$ years the particle temperatures are in the chondrule formation range. For these calculations, we have used standard values for the solar, pre-main sequence rotational period (8 days) (Mathieu 2004), magnetic field strength (0.1 T) (Güdel 2007) and mass accretion rates.

It should be noted that these temperatures are dependent on the assumed magnetic field strength of the star. If the field strength was greater than 0.1 Tesla then $R_t$ would increase and the particle temperature decrease. The reverse is true if the field strength was lower than 0.1 Tesla.

3. Chondrule Oxygen Isotopic Reservoirs

Possible evidence against the Jet Flow Model is the distinct oxygen isotopic reservoirs observed in chondrules from different classes of chondrites (Scott 2007). This raises the possibility that chondrules were formed by localized heating events, such as shock waves, at different locations within the solar nebula. Such distinct oxygen reservoirs may, however, be due to temporal rather than spatial differences in the solar nebula.

By virtue of its location, the inner edge of the solar nebula was most likely the major mixing region between the $^{16}$O-rich solar wind (Nakamura et al. 2008) and the $^{16}$O-poor solar nebula (Yurimoto & Kuramoto 2007). This would be particularly true for condensates (e.g, CAIs and AOAs) that formed in the jet flow. Such jet flow condensates would tend to be richer in $^{16}$O relative to the solar nebula, because of the direct interaction of the solar wind with the gases in the jet. The oxygen isotopic composition of melt objects such as chondrules would reflect the average oxygen composition of dust (e.g., original solar nebula dust plus previously processed jet flow materials) and gas at the inner solar nebula. Thus, one would expect a mixing line with a slope 1 between the $^{16}$O-rich jet flow condensates and the $^{16}$O-poor melt products (Figure 4). This "jet flow mixing line" would be offset from the standard CCAM (carbonaceous chondrite anhydrous mineral) line which has a slope of ~ 0.96 (Young & Russell 1998).

As such, the observed oxygen isotopic composition of chondrules may indirectly reflect the density of solar wind at the inner edge of the solar nebula. During times of low solar wind density, any chondrules produced by the flow would have oxygen isotope values close to the solar nebula value at the inner edge of the disk. If the strength of the jet flow was somewhat proportional to the density of solar wind at $R_t$ then the jet flow would also be relatively weak and would not loft the chondrules all that far and so we should expect a bias for $^{16}$O-poor chondrules in the inner nebula.

In the theory for jet flow formation outlined in Liffman (2007), the strength of the wind flow is low when $R_t \approx R_{co}$. From the results shown in Figure 3, the formation of the inner nebula chondrules may have occurred at around $10^6$ years when $R_t$ was slightly less than $R_{co}$. Assuming an approximately spherically symmetric and average value for the solar wind, the number density, $n$, of the average solar wind at $R_t$ would be

$$n(R_t) \approx n(R_*)\left(\frac{R_*}{R_t}\right)^2. \tag{5}$$

So there should be a correlation between solar wind density, $^{16}$O enrichment and the value of $R_t$. If $R_t$ is small then the solar wind density is large, as is the subsequent $^{16}$O enrichment. If $R_t$ is large then the solar wind density and $^{16}$O enrichment is smaller.

During times of relatively high solar density at $R_t$, the condensates and subsequent chondrules formed in the jet flow would have been relatively enriched in $^{16}$O. If high solar wind intensity at the inner edge of the nebula is also correlated with stronger outflows then there would be a tendency for $^{16}$O-rich chondrules to have been transported to outer regions of the nebula. From Figure 3 and equation (5), such a situation may have occurred when $R_t$ was relatively close to the Sun, i.e., $R_t$ was significantly smaller than $R_{co}$ and the age of the nebula was greater than $10^6$ years. This is also consistent with the results found in Liffman (2007), where the strength of the jet flow increases as the inner edge of the nebula approaches the stellar surface.

From this reasoning, one would expect older, $^{16}$O-poor chondrules would be more likely to be present in the inner regions of the nebula, while younger, $^{16}$O-rich chondrules would tend to be found in the outer regions of the nebula.

Two additional results arise from this theory: (i) the solar nebula would have become more $^{16}$O-rich with time, (ii) there would have been a negative gradient of $^{16}$O in the solar nebula, i.e., the relative amounts of $^{16}$O would have decreased with increasing radial distance, $r$, from the Sun. The first effect arises, because the jet flow was producing relatively $^{16}$O-rich solid material and some of this material was falling back to the solar nebula, so the amount of $^{16}$O-rich material in the nebula would have increased with time. The second effect is due to the disc shape of the nebula, where the density of in-falling material from the jet flow would have decreased with increasing distance from the Sun, because the area of the nebula increases as $r^2$.

Figure 5 summarizes some of these ideas with a schematic representation of the oxygen isotopic evolution of both the solar nebula and the chondrules produced in the solar bipolar jet flow. It is important to note that the inner edge of the solar nebula (denoted by SN) becomes more $^{16}$O-rich with time. The inner nebula chondrules also follow this trend, i.e., younger chondrules are more enriched in $^{16}$O. In particular, from the oxygen isotopic values of chondrules as given in Scott (2007), this model would predict that R chondrules are older than ordinary chondrite chondrules which are older than enstatite chondrules.

At around a million years after the formation of the solar nebula, the formation of inner nebula chondrules starts to switch off and the jet flow starts producing outer nebula chondrules. For these chondrules, the $^{16}$O enrichment of chondrules is dependent on the density of the solar wind at $R_t$, as well as the steady $^{16}$O enrichment at $R_t$, so the system is more variable and the correlation between youth and $^{16}$O enrichment is not clear. None-the-less, the model outlined here, suggests that inner nebula chondrules (R, OC & E) should be older than the outer nebula chondrules (e.g., chondrules from carbonaceous chondrites).

It is observed, that many chondrules, CAIs and other related meteoritic objects have oxygen isotope ratios that do not lie on the slope 1, Jet Flow Mixing Line. This is probably due to subsequent parent body processing, which produces mass-dependent fractionation and acts to spread the 3 isotope values on lines parallel to the Terrestrial Fractionation (TF) line (Young & Russell 1998).

## 4. Chondrule Types

In any individual unequilibrated chondrite, there are low-FeO (Type I) porphyritic chondrules and high-FeO (Type II) porphyritic chondrules that have different O isotopic compositions. The Type I chondrules tend to be $^{16}$O rich and the Type II, $^{16}$O poor. It seems likely that the Type I chondrules formed earlier than the Type II chondrules, since they sometimes occur as relicts in the Type-II chondrules (Jones et al (2000), Kunihiro et al. (2004, 2005)).

In the Jet Flow Model, chondrule type is probably dependent on the value of $R_t$. For example, suppose there was a sudden surge in accretion (an occurrence which is consistent with observations of accretion in young stars (Hartmann 1998, Frank et al. 2002)). This would increase $\dot{M}_a$ and would, by equation (1), decrease $R_t$, moving the inner edge of the nebula closer to the Sun. From equation (4), the temperature at $R_t$ would increase as would the density of the solar wind and the relative amount of $^{16}O$ (eq. (5)). It is at this stage that the low-FeO, Type I chondrules may have formed.

If the rate of accretion started to decrease then, $R_t$ would increase, the temperature at $R_t$ would decrease as would the density of the solar wind and the $^{16}O$ enrichment. This would be the formation stage for the high-FeO Type II chondrules. Such a scenario suggests that Type I chondrules have, on average, a higher formation temperature relative to Type II chondrules - a result which is consistent with observations (Hewins and Radomsky 1990).

This theory suggests that if we could date both the Type I relic grains and their encapsulating Type II chondrules then we may obtain a time scale for a single accretion event in the solar nebula. Indeed, it is possible that chondrules are a temporal record of solar nebula accretion events.

## 5. Discussion

The temperature of the inner disk edge is due to the interplay between mass accretion onto the star plus the luminosity and radius of the star. As the luminosity of the star and the mass accretion rate onto the star decreases, the temperature of the inner disk decreases. The decreasing mass accretion also causes the inner rim to move away from the star. However, at around $10^5$ years, the decreasing radius of the star decreases the stellar magnetic field strength and the inner rim starts moving in towards the star. At this stage, the inner rim temperature increases. Eventually, at around $10^6$ years, the luminosity of the star decreases sufficiently for the inner disk rim temperature to decrease over a time scale of millions of years.

The position of the inner edge of the solar nebula relative to the co-rotation radius determines the power of the solar bipolar jet flow and where the jet flow products eventually land in the solar nebula. The position of the inner solar nebula edge relative to the Sun also determines the formation temperature and the oxygen isotopic composition of jet flow products. It is the relative movement of the inner solar nebula edge with time that produces the different isotopic reservoirs observed in CAIs, AOAs, Type I and II chondrules and other, similar meteoritic components.

As such, there is no need to invoke shock heating to produce localized high-temperature zones in the solar nebula. The inner edge of the solar nebula appears to have had the appropriate temperatures, $^{16}O$ enrichments and timescales to produce most of the observed high-temperature particles found in comets and meteorites.

Acknowledgements

The author wishes to acknowledge the constructive, good-natured, objective comments and questions from Alan Rubin.

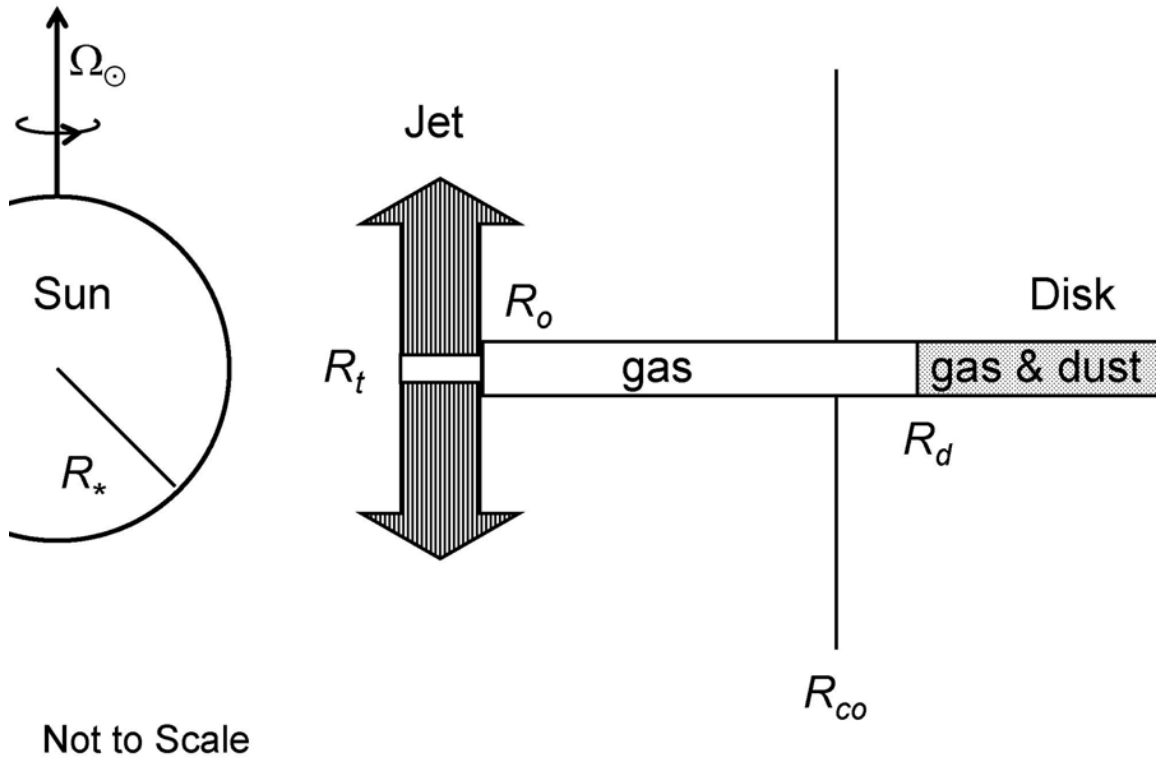

Figure 1: Structure of the inner solar nebula with $R_t$ - disk truncation radius, $R_*$ - radius of a star/Sun, $R_o$ – outer radius of the jet flow, $R_{co}$ - co-rotation radius, $R_d$ - dust sublimation radius and $\Omega_\odot$ - angular rotational frequency of the Sun. Particles are formed at $R_t$ and ejected over the disk. For illustrative purposes, the case of $R_t < R_{co}$ is shown.

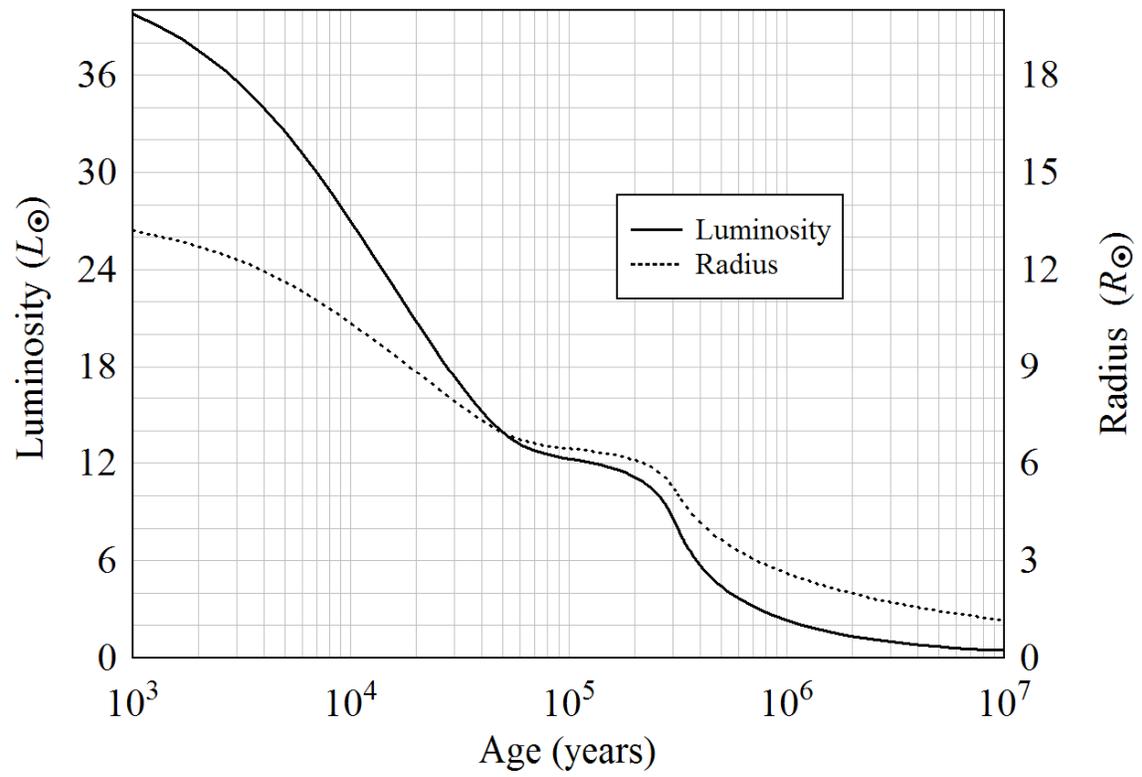

Figure 2: Luminosity and radius of a Sun-like star (X = 0.703, Y = 0.277, Z = 0.020) for the first 10 million years (Siess et al. 2000).

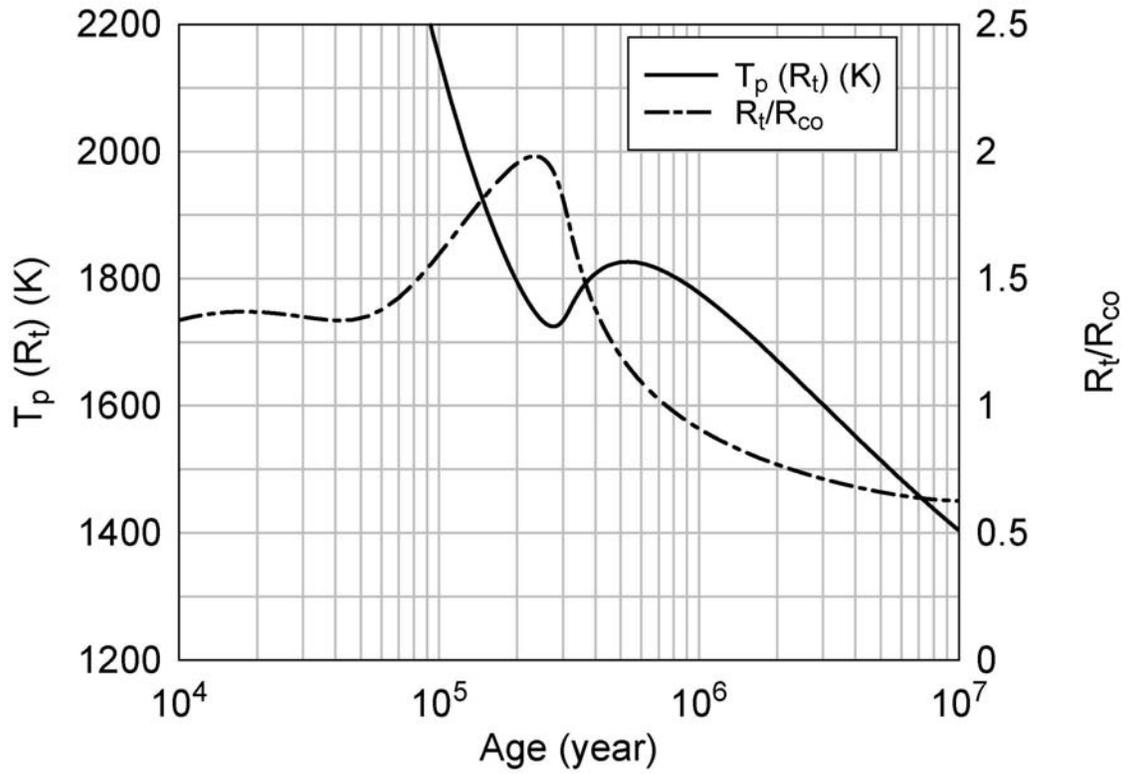

Figure 3: Particle temperature, $T_p$, at the inner edge of the solar nebula and $R_t/R_{co}$ as a function of time.

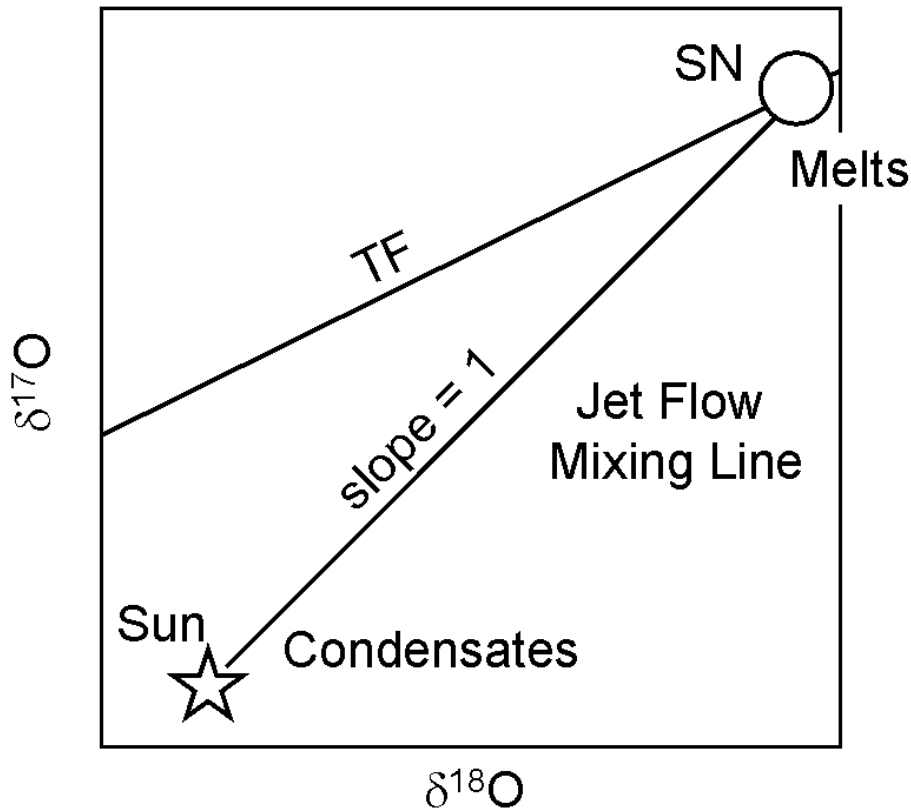

Figure 4: Material produced in a jet flow has oxygen isotopic ratios that lie on a mixing line (of slope 1) connecting the solar wind (represented by the Sun symbol) and the inner rim of the solar nebula (SN). The oxygen isotopic values for jet flow condensates (e.g., CAIs) and melts (e.g., chondrules) lie on this mixing line with the condensates close to the solar wind value and the melts close to the solar nebula value. TF – terrestrial fractionation line, $\delta^{17}O$ and $\delta^{18}O$ are the permil deviation $^{17}O$ and $^{18}O$ relative to $^{16}O$ ( i.e., $\delta^{i}O = 10^{3}((^{i}O/^{16}O)/(^{i}O/^{16}O)_{SMOW}-1)$, where SMOW refers to Standard Mean Ocean Water.

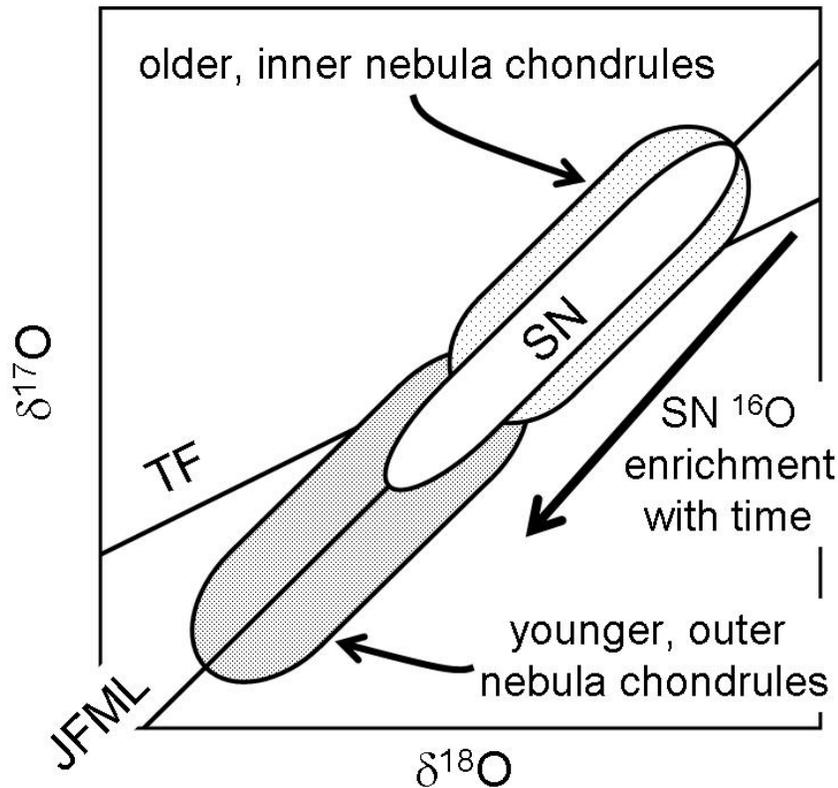

Figure 5: Oxygen isotopes of chondrules and the inner edge of the solar nebula (SN). Due to infall from jet flow, the solar nebula becomes more $^{16}$O-rich with time. The Jet Flow Mixing Line (JFML) is the (slope 1) oxygen isotopic mixing line between the solar wind and the inner edge of the solar nebula. The relatively $^{16}$O-poor chondrules are formed at $t \sim 10^6$ years and are projected into the inner nebula via a relatively weak outflow. The relatively $^{16}$O-rich chondrules are formed at $t > 10^6$ years and are projected to the outer nebula by a relatively strong outflow. Not shown is the negative $^{16}$O enrichment gradient in the solar nebula as a function of distance from the Sun.